# Picosecond laser source at 1 MHz with continuous tunability in the visible red band

Sébastien Forget, François Balembois, Gaëlle Lucas-Leclin and Patrick Georges

*Laboratoire Charles Fabry de l'Institut d'Optique, UMR 8501 du CNRS*
*Université Paris-Sud*
*91403 Orsay cedex*

**Abstract** :

We report the first demonstration to our knowledge of a continuously tunable picosecond laser operating around 1 MHz. The emission can be tuned from 640 to 685 nm and the repetition rate from 200 kHz to 1 MHz with a pulse duration of less than 200 ps. The system is based on a Nd:YVO$_4$ passively Q-switched microchip laser providing a few tens of nJ per pulse. Two cascaded stages of amplification are then used to increase the pulse energy to several µJ. The frequency doubled radiation is then used to pump a periodically-poled-niobate-lithium (PPLN)-based optical parametric generator in an all-solid-state architecture. 20 nJ of tunable signal radiation are obtained. We also demonstrated 300-ps pulses generation in the UV (355 nm) at 1 MHz.

Many applications in biology such as fluorescence lifetime measurements, detection of single molecules or optical tomography by time correlated single photon counting require pulsed laser sources emitting visible radiations. As the fluorescence lifetime of numerous molecules is less than a nanosecond, short pulses (around one hundred of picosecond) are needed. The repetition rate of the laser source used is an important parameter in experiments using photon counting detection chains : indeed, high repetition rates ensure a fast acquisition process of the fluorescence decay signal and allow to study of dynamic processes. However, the complex signal processing devices could not support repetition rates above several MHz without problems during the acquisition of the data. Consequently a repetition rate of 1 MHz seems to be a good requirement for this kind of applications. Another very important characteristic is the tunability of the source: in order to study different chemical species with the same laser source, it is very interesting to be able to change the wavelength easily on a broad range. There is consequently a strong need for tunable sources operating at 1 MHz in a compact and simple package.
A fruitful approach to generate short visible pulses is to use a master oscillator power amplifier system operating in the near-infrared and to use non-linear processes to reach the visible range. Recent developments in passively Q-switched microchip lasers [1,2] open the way to a large number of studies on very compact and efficient pulsed sources [3-7]. Nevertheless, the repetition rates of the published devices are almost always in the kilohertz range. Moreover the poor efficiency of non linear materials at weak peak powers makes the development of MHz sources somewhat difficult. An interesting and successful approach is to use a fiber-based architecture and quasi-phase matched non-linear materials [8].
In this paper we present another solution based essentially on bulk elements to demonstrate a compact source providing 45 nm of continuous tunability around 660 nm with less than 200 ps long pulses. To this end, we used a Nd:YVO$_4$ passively Q-switched amplified microchip laser operating at 1 MHz to pump a periodically poled lithium niobate optical parametric generator. We also demonstrate 50 mW of 355 nm radiation by frequency tripling the same microlaser.

Motivated by compactness and simplicity we chose as master oscillator a microchip Nd:YVO$_4$ laser. We used a semiconductor saturable absorber mirror to passively Q-switch the laser [9]. The repetition rate range and the duration of the pulses could then be adjusted by choosing the proper parameters for the SESAM and the design of the cavity.
It can be shown that the pulse duration is directly proportional to the length of the cavity, which led us to design a very short cavity (a few hundreds micrometers) to get sub-nanosecond pulses. We used a 200 µm thick laser crystal with an antireflection coating at 1064 nm on its faces. The saturable absorber had a modulation depth of 6% and we used several output couplers between 4 and 10 %. The crystal used was a 3% doped Nd:YVO$_4$ and was sandwiched between the saturable absorber and the output coupler as described on figure 1.

To improve the compactness of the system, we use the same laser diode to pump the microchip laser and the power amplifier : the beam of a 200 µm-wide 20 W fiber-coupled laser diode at 808 nm was divided into two parts : about 1 W was taken to pump the microchip laser via a small mirror placed in the path of the laser diode beam as shown on figure 3, while the rest of the laser diode beam (19 W) was used to pump the power amplifier. This setup provided simplicity, compactness and reducing the global cost. It also allows us to use a high brightness pump beam for the microchip laser : to give an order of magnitude, the microchip pump beam is equivalent in terms of brightness to the beam coming from a fiber coupled laser diode providing 9 W with a numerical aperture of 0.2 and an emitting area of 200 µm in diameter, which is more than 4 times as bright as the pump diodes generally used for this kind of devices [2]. The radius of the pump spot on the microchip was around 30 µm.

We obtained single-frequency Q-switched sub-nanosecond pulses and a repetition rate around 1 MHz for 200 mW of incident pump power. The repetition rate can be easily tuned from 200 kHz to 2 MHz by controlling the incident pump power with an iris whereas the pulse duration remained nearly constant as expected [1] : about 900 ps for the 4% output coupler and 400 ps for the 10% coupler.
The figure 2a summarizes the pulse durations and repetition rates obtained with two different output couplers. The pulse-to-pulse timing jitter is typically 2% and the beam was not diffraction limited ($M^2 = 2.2$). We got a pulse energy of only 20 nJ and 60 nJ with the 4% and 10% output couplers respectively (figure 2b). The efficiency of non-linear processes is related to the peak power, and such energies correspond to obviously too low peak powers – about 100 W - for efficient frequency conversion. Amplification stages are therefore needed to scale this peak power towards the multi-kilowatt level.
As our goal was to reach very short pulses (some hundreds of picoseconds maximum) with a repetition rate of 1 MHz, we chose the 10% output coupler for the following experiments.
The 3D multipass amplifier used in our system was described in details in ref. 10. The laser beam made 6 passes (only 4 are represented on figure 3) in a 10 mm-long 0.1% doped Nd:YVO$_4$ crystal in three different planes of incidence to optimise the overlap between the laser beam path and the pumped volume. The use of both low-doped and relatively long crystal allows a better distribution of the pump power inside the gain medium, leading to a weaker and less aberrating thermal lens when compared with more doped crystals. It also authorises a higher pumping level (up to 20 W in our case) without fracture problems.

The power amplifier is pumped with the laser diode described below (see figure 2). With an average power of 40 mW from the microchip laser injected into the 3D multipass amplifier, we obtained 2.8 W (corresponding to a peak power of about 5.6 kW) at the output of the amplifier. The repetition rate, the pulse length and the pulse-to-pulse stability were preserved after amplification.
However the spatial beam quality is again slightly degraded in the amplifier, leading to likely poor frequency conversion efficiency. Moreover, at this level of input power all the energy stored in the amplifying crystal was not extracted in the 3D amplifier : in order to improve the amplification, and to spatially filter in the same time the microchip laser beam, a commercial (Keopsys, model KPS-GB-YFA-27-FA) ytterbium doped fiber preamplifier was placed between the microchip laser and the multipass amplifier. This preamplifier worked in a saturated regime, providing more than 100 mW of output average power for a large range of input powers from 100 µW to several milliwatts. The output average power obtained after the preamplifier-amplifier system was 4 W of 1064 nm radiation. Another advantage of the YDFA preamplifier is that this output powers remained constant whatever repetition rate was used, and the beam was now quasi-diffraction limited ($M^2 = 1.2$).
We used a periodically poled KTP to perform single-pass second harmonic generation from 1064 to 532 nm. The PPKTP was 10 mm long and placed in an oven to control the temperature. The 1064 nm laser beam is focused by a 150 mm lens to a waist radius of 50 µm. For 3.7 W of incident power we obtained 1.4 W at 532 nm (38 % of efficiency). The pulse duration was slightly reduced during the frequency doubling process (from 500 ps to 400 ps, see figure 5).

We then used this 532 nm radiation to pump an optical parametric generator (OPG) based on a quasi-phase-matched crystal in order to perform continuous tunability in the red part of the spectrum around 660 nm. In this purpose we chose a periodically-poled lithium niobate crystal (PPLN) because of its exceptionally high effective non-linear coefficient of 17 pm/V. Such PPLN-based OPG configurations had already been successfully investigated in the ns regime with low repetition rates [5-7].
The PPLN crystal was 3 cm-long and consisted in 5 parallel gratings. The five periods were 9, 9.25, 9.5, 9.75 and 10 µm, each with a width of 1 mm and a height of 0.5 mm. The end faces of the crystal were uncoated, giving rise to 14 % Fresnel reflection losses from each face. The crystal was placed in an oven to avoid photorefractive damage and to control precisely the temperature. The nominal temperature was 90 °C to obtain phase-matching at 660 nm (the idler being 2743 nm in this case) for the 9.5 µm-period grating.

The 532-nm pump laser was focused at the center of the non-linear crystal to a 50 µm waist radius by a 150-mm lens (figure 3).. The oven containing the PPLN could be easily pushed sideways via a micrometric screw to choose one of the five gratings.

The temperature of the oven could also be tuned from 30°C to 150 °C to complete continuous tunability for the signal from 640 to 685 nm (3.15 µm to 2.38 µm for the idler) as shown on figure 4. A slight discrepancy between the experimental data and the theoretical fit based on the published Sellmeier equation [11] is observed. Some imprecision could affect the values : errors in the knowledge of the Sellmeier law giving the different indexes, on the exact temperature and on the period of the gratings. We estimated a global potential error on the signal wavelength around 1.1 nm, which makes the experimental data and theoretical results consistent. However, a more detailed study should be necessary to explain the slight slope difference between experiment and theory.

With 1.2 W of pump power at a repetition rate of 1 MHz and with a pulse duration of 400 ps (corresponding to a peak power of 2kW), the signal average power was only 15 mW. The OPG operated in fact just above its threshold which was found to be 1.4 kW of peak power. The low efficiency could be explained by the not-well-understood effects that affect the PPLN when pumped in the visible at relatively high average powers [12-14]. Indeed, we previously performed with our PPLN an OPG experiment using a commercial (DualChip by JDS Uniphase) frequency-doubled microchip laser [15]. In that case the repetition rate was only 40 kHz and the average power around 100 mW, leading to comparable peak powers. The results obtained were consistent with the better comparable published results, confirming the relatively good quality of the poling of our PPLN sample. In the setup operating at 1 MHz, the much higher average power induced beam distortion and even local bulk damage, consequently leading to a decrease of the efficiency. The pulse duration of the signal is strongly reduced to less than 200 ps (limited by the rise time of our measurement setup) as shown on figure 5, because of the OPG exponential gain expression in the high-gain regime.

Those results were obtained with the 9.5-µm grating, but the thresholds and efficiencies were the same for the five different gratings, since the idler's wavelength remained always under 4 µm were the lithium niobate absorbs laser energies.

A large number of biological molecules have to be excited by ultraviolet pulses, which induces a great interest of UV laser sources which short pulses durations (typically 100 ps) and high repetition rate. To efficiently frequency double the red radiation, we could use UV-transparent quasi phase matched crystals as periodically poled lithium tantalate, but such crystals are not easily grown [12] because of the short period required and at this time not commercially available. However we performed frequency tripling of our amplified microchip laser system with a bulk LBO crystal to obtain a 355 nm UV radiation. We separate the green and infrared beams in order to rotate the 1064 nm polarisation of 90 degrees to get type II phase matching in a 7 mm-long LBO crystal. With 1.1 W of 532 nm and 1.6 W of 1064 nm radiation we obtained only 50 mW of ultraviolet radiation because of unoptimized design. We haven't measure the temporal width of the ultraviolet pulses because our photodiodes are not sensitive enough in this range of wavelength, however theoretical considerations shows that the UV width should be around 300 ps. Great improvement in terms of efficiency and tunability could be obtained by the use of periodically poled lithium tantalate to perform THG and to frequency double the output of the OPG respectively. The later point will allow us to get a picosecond laser source with large tunability around 330 nm at high repetition rate, which is of great interest for many biological studies.

In conclusion, we demonstrate what is to our knowledge the first compact diode-pumped sub-nanosecond source providing continuous tunability over 45 nm in the red range of the spectrum (around 660 nm) with a high repetition rate of 1 MHz. This source could be frequency doubled to reach tunability around 330 nm with the same temporal characteristics. We also get 300 ps pulses at 355 nm with the same repetition rate and expect to reach in a near future.

**Acknowledgements**
*This work has been partially supported by the research program POLA from the Contrat Plan Etat Région (2000-2006) (French State and Conseil Général de l'Essonne).*

**References**

1. J.J. Zayhowski and C. Dill III, Opt. Lett. 19 (18), 1427-1429 (1994).
2. B. Braun, F.X. Kärtner, U. Keller, J.-P Meyn and G. Huber, Opt. Lett. 21 (6), 405-407 (1996).
3. J.J. Zayhowski, Opt. Lett. 21 (8), 588-590 (1996).
4. F.Druon, F.Balembois, P.Georges and A.Brun, Opt. Lett. 24, 499-501 (1999).
5. A.C. Chiang, Y.C. Huang, Y.W. Fang and Y.H. Chen, Opt.Lett 26 (2), 66-68 (2001)
6. K.W Aniolek, R.L Schmitt, T.J. Kulp, B.A Richman, S.E Bisson and P.E Powers, Opt. Lett. 25, 557-559 (2000).
7. U.Bäder, J.P Meyn, J.Bartschke, T.Weber, A. Borsutzky, R. Wallenstein, R.G Batchko, M.M Fejer and R.L Byer, Opt. Lett. 24, 1608-1610 (1999).
8. P.A. Champert, S. Popov, M. Solodyankin and J.R. Taylor, Elect. Lett, 38 (13) 627-628 (2002).
9. B. Braun, F.X. Kärtner, G. Zhang, M. Moser and U. Keller, Opt. Lett. 22 (6), 381-383 (1997).
10. S. Forget, F. Balembois, P.J Devilder and P. Georges, Appl. Phys. B. 75, 481-485 (2002)
11. D. Jundt, Opt. Lett. 22, 1553-1555 (1997).
12. G.D Miller, R.G Batchko, W.M Tulloch, D.R Weist, M.M Fejer and R.L Byer, Opt.Lett 22 (24) 1834-1836 (1997)
13. S.C. Tidwell, J.Seamans and P.Roper, paper CPD25 in Conf. on Lasers and Electro-Optics (CLEO '97), 1997 OSA Technical Digest Series (Washington DC: Optical Society of America)
14. R.G Batchko, G.D Miller, A Alexandrovski, M.M Fejer and R.L Byer, Paper CTuD6 in Conf. on Lasers and Electro-Optics (CLEO '98), 1998 OSA Technical Digest Series (Washington DC: Optical Society of America)
15. E.Herault, S.Forget, G. Lucas-leclin and P. Georges, in OSA Trends in Optics and Photonics (TOPS) vol 69, Advanced Solid State Photonics, OSA Technical digest Post Conference Edition (Optical Society of America, Washington DC, 2003) pp 313-315.
16. J.-P. Meyn, C.Laue, R.Knappe, R.Wallenstein, M.M. Fejer, Appl. Phys. B 73, 111-114 (2001)

# Figures

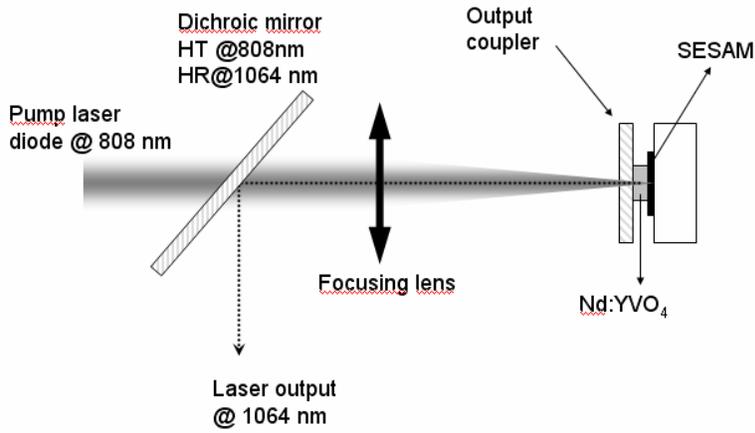

Figure 1

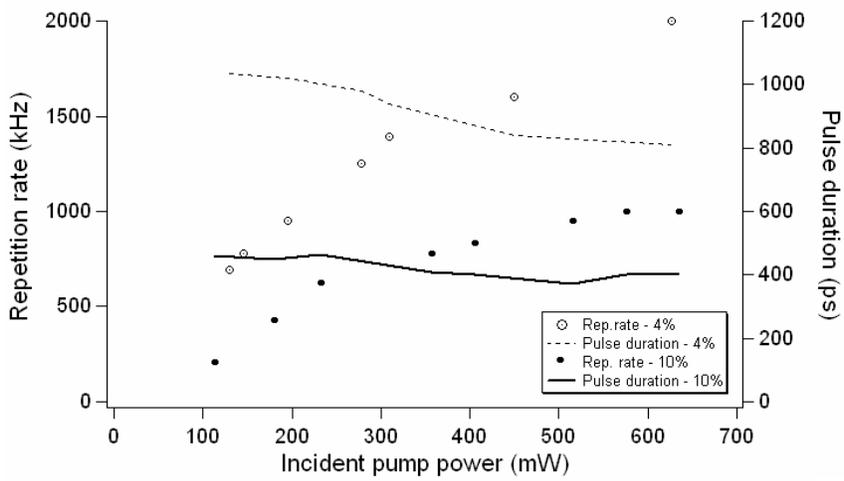

Figure 2a

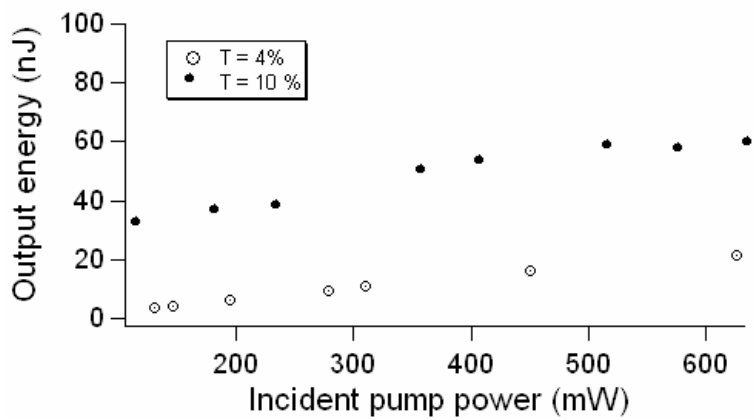

Figure 2b

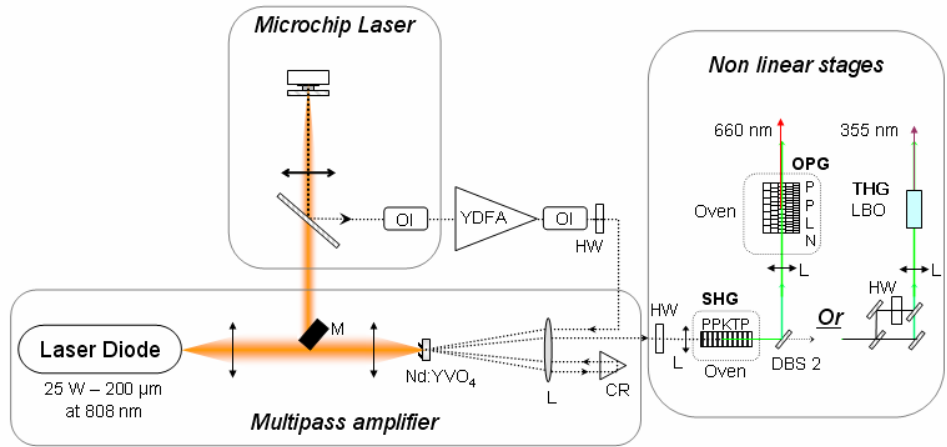

Figure 3

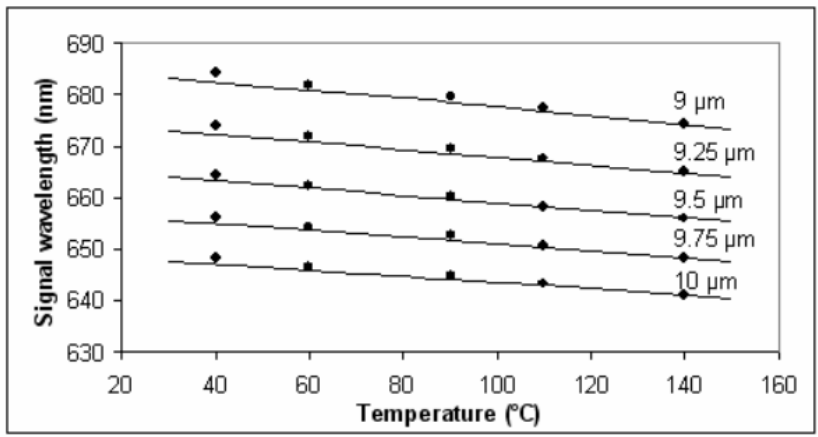

Figure 4

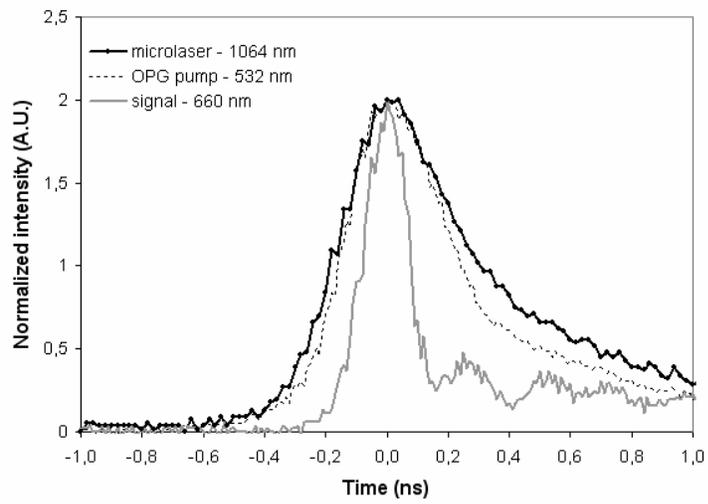

Figure 5

# **Figure captions**

*Figure 1 : Layout of the Q-switches Nd:YVO$_4$ microchip laser with 200µm-wide cavity*

*Figure 2: (a) temporal characteristics of the microchip laser (repetition rate and pulse duration) relatively to the incident pump power for two different output couplers (T = 4 and 10 %) - (b) Output pulse energy from the microchip laser relatively to the incident pump power for both output couplers*

*Figure 3 : Experimental setup. The 3D multipass amplifier presented here is a 4-pass to simplify the drawing. 6-pass configuration is obtained simply by adding another cornercube retroreflector (CR). OI are the optical isolators used to protect the YDFA and to prevent parasitic laser oscillation , HW are 1064 nm half-wave plates, and DBS are dichroïc beam splitters (HR at 532 nm, HT at 1064 nm).The two different non-linear stages are presented (see text).*

*Figure 4 : Tuning range of the OPG with the temperature for the 5 gratings. Dots correspond to experimental measurements. The solide curve represents the signal wavelength calculated with the Sellmeier equation given in Ref. 11.*

*Figure 5 : pulse durations for the amplified microchip laser (FWHM 510 ps), the green pump of the OPG (400 ps FWHM) and the red signal (less than 200 ps FWHM)*